\documentstyle[psfig,12pt]{article}

        \newlength{\paperbaselineskip}
\newfont{\fourteencp}{cmcsc10 scaled\magstep2}
\newfont{\titlefont}{cmbx10 scaled\magstep3}
\newfont{\authorfont}{cmcsc10 scaled\magstep1}
\newfont{\fourteenmib}{cmmib10 scaled\magstep2}
        \skewchar\fourteenmib='177
\newfont{\elevenmib}{cmmib10 scaled\magstephalf}
        \skewchar\elevenmib='177
\newfont{\ninemib}{cmmib9} \skewchar\ninemib='177
\makeatletter
\newif\ifpr@pstyle \pr@pstylefalse
\newif\ifnons@qeq  \nons@qeqfalse
\newcommand\nonsequentialeqnum{
        \nons@qeqtrue
        \@addtoreset{equation}{section}
        \def\theequation{\arabic{section}.\arabic{equation}}}
\newif\ifp@bblock  \p@bblocktrue
\newcommand\nopubblock{\p@bblockfalse}
\newcommand\topspace{\hrule height 0pt depth 0pt \vskip}
\newcommand\p@bblock{\begingroup \tabskip=\hsize minus \hsize
        \baselineskip=1.5\ht\strutbox \topspace-2\baselineskip
        \halign to\hsize{\strut ##\hfil\tabskip=0pt\crcr
        \the\Pubnum\crcr\the\date\crcr}\endgroup}
\newcommand\YUKAWAmark{\hbox{
        \ifpr@pstyle\ninemib\else\elevenmib\fi
        Yukawa\hskip1mm Institute\hskip1mm Kyoto \hfill}}
\newtoks\date
\newtoks\Pubnum

\newcommand{\frontpageskip}{\vspace{12pt plus .5fil minus 2pt}}
\def\@authoraddress{} \def\@title{}
\def\title#1{\gdef\@title{\frontpageskip
        \begin{center}{\titlefont #1}\end{center}\par}}
\def\@author#1{\frontpageskip\par\begin{center}{\authorfont #1}
        \end{center}
        \nobreak}
\def\author#1{\expandafter\def\expandafter\@authoraddress\expandafter
    {\@authoraddress{\@author{#1}}}}
\def\andauthor#1{\expandafter\def\expandafter\@authoraddress\expandafter
    {\@authoraddress{\frontpageskip\centerline{and}\@author{#1}}}}
\def\authors#1{\expandafter\def\expandafter\@authoraddress\expandafter
    {\@authoraddress{\frontpageskip\noindent #1}}}
\def\@address#1{\par\begin{center}{\sl #1}\end{center}\par}
\def\address#1{\expandafter\def\expandafter\@authoraddress\expandafter
    {\@authoraddress{\@address{#1}}}}
\def\andaddress#1{\expandafter\def\expandafter%
    \@authoraddress\expandafter
    {\@authoraddress{\par\centerline{\sl and}\@address{#1}}}}
\renewcommand{\thanks}[1]{\footnote{#1}}
\def\maketitle{\par
  \begingroup
       \def\thefootnote{\fnsymbol{footnote}}
        \thispagestyle{empty}
        \baselineskip=\paperbaselineskip
        \@maketitle
        \endgroup
        \setcounter{footnote}{0}
        \let\maketitle\relax \let\@maketitle\relax
        \let\@thanks\relax \let\@title\relax
        \let\@title\relax \let\@authoraddress\relax
        \let\thanks\relax}
\def\@maketitle{%
        \ifpr@pstyle\vspace{-1.0cm}\else\vspace{-1.7cm}\fi
        \YUKAWAmark\vskip0.6cm
        \ifp@bblock\p@bblock \else\hrule height 0pt \relax \fi
        \@title
        \@authoraddress
        }
\renewcommand{\abstract}{\par\vspace{80pt plus 3pt minus 3pt}
        \frontpageskip\centerline{
             \ifpr@pstyle\twelvecp\else\fourteencp\fi Abstract}
        \vspace{8pt plus 3pt minus 3pt}}
\makeatother
\def\comptblwd#1#2{%
\expandafter\gdef\csname tref@#1\endcsname{#2}%
}

\def\m@thcombine#1#2{%
  \setbox0=\hbox{$#1$}
  \setbox1=\hbox{$#2$}
  \ifdim\wd0>\wd1
    \setbox0=\hbox to\wd1{\hss\box0\hss}
  \else
    \setbox1=\hbox to\wd0{\hss\box1\hss}
  \fi
  \mathop{\vcenter{
    \offinterlineskip\box0\box1}}}
\def\lesim{\m@thcombine<\sim}
\def\gesim{\m@thcombine>\sim}

\newcommand{\ket}[1]{| {#1} \rangle}
\newcommand{\bra}[1]{\langle {#1} |}
\newcommand{\hket}[1]{ {#1} \rangle}

\begin{document}
\Pubnum={YITP-99-36}
\date={June 1999}


\title{Probing the width of compound states \break 
with rotational gamma rays}

\author {M.~Matsuo$^{1}$,
T.~D\o ssing$^{2}$, B.~Herskind$^{2}$, 
S.~Leoni$^{3}$, E.~Vigezzi$^{3}$, R.A.~Broglia$^{2,3}$}

\address{$^1$ Yukawa Institute for Theoretical Physics, 
Kyoto University, Kyoto 606-8502, Japan}

\address{$^2$ Niels Bohr Institute, University of Copenhagen,
DK2100 Copenhagen \O, Denmark}

\address{$^3$ INFN sez Milano, and Department of Physics, 
University of Milano, Milan 20133, Italy}

\maketitle

\begin{abstract}
The intrinsic width of (multiparticle-multihole) compound states is an
elusive quantity, of difficult direct access, as it is masked by damping
mechanisms which control the collective response of nuclei. Through
microscopic cranked shell model calculations, it is found that the
strength function associated with two-dimensional gamma-coincidence
spectra arising from rotational transitions between states lying at
energies $>$ 1 MeV above the yrast line, exhibits a two-component structure
controlled by the rotational (wide component) and compound (narrow
component) damping width. This last component is found to be directly
related to the width of the multiparticle-multihole autocorrelation
function.
\vspace{15mm}



\end{abstract}


\vspace{10mm}

In deformed nuclei, the observed discrete rotational bands are often
successfully described as states of a cranked mean field \cite{Aberg1}.
For fixed angular momentum and
increasing excitation energy,
the residual interaction not included in the mean
field will eventually generate compound states, which are
superpositions of the many-particle many-hole mean field states. As a
result, each basis band state $\ket{\mu}$ becomes distributed over
the compound states $\ket{\alpha}$
within an energy interval known as the compound
state damping width $\Gamma_{\mu}$ \cite{BM,Lauri86,Rev}.

The quantity $\Gamma_{\mu}$ plays a central role in the study of 
basic nuclear phenomena,  like the statistical and chaotic
features of energy levels \cite{Mottelson,Zelevinsky,Persson-Aberg}, 
or the damping of collective vibrations \cite{Lauri88,Lauri95}.
However, it also appears to be inaccessible
by direct experimental means, since it is essentially not possible to
excite a pure 
many-particle many-hole state. We shall demonstrate that the 
spectrum of collective E2-gamma rays emitted by the compound states
built out of rotational bands carries information about $\Gamma_{\mu}$.
This is true also for the unresolved gamma rays,
which are far too weak to allow
for construction of a level scheme with present experimental techniques. 

Although rotational damping is a phenomenon which is independent of compound 
damping, 
being controlled by fluctuations in the alignment of the single-particle
states, the occurence of compound states in rotational nuclei 
is usually accompanied by damping of rotational motion
\cite{Lauri86,Rev,Leander,FAM,RPM}. 
In what follows we shall study the interplay
between these two independent phenomena, namely rotational damping and
compound damping, as a function of spin and excitation energy, making use
of a cranked shell model  
which has been applied earlier to the study of rotational
damping and of the statistical properties of spectral fluctuations
and level distances \cite{Aberg2,Matsuo93,Matsuo97}.
The calculations
have been performed for the rare-earth nucleus $^{168}$Yb, for which 
the quasi-continuum gamma spectrum has been analyzed in detail experimentally.
The shell model Hamiltonian, consisting of the cranked Nilsson
mean-field and the surface-delta interaction acting as the residual
two-body force, is diagonalized using 
the lowest 2000 many-particle many-hole configurations based on
the cranked Nilsson single-particle orbits
for each value of average angular momentum $I$ and the parity $\pi$.
This provides the lowest 600 energy levels for each $I^{\pi}$
covering an energy range up to about 2.5 MeV above the
yrast line. (See ref. \cite{Matsuo97} for further details).
In the calculation, rotational damping sets in at about 1 MeV above the
yrast line (in agreement with  experiments) as a consequence
of the spreading of the unperturbed rotational bands 
having specific and simple shell model
configurations in a rotating deformed mean-field.
Above the onset energy and up to a few MeV, two-particle two-hole
(2p2h) and three-particle three-hole (3p3h) configurations
are the dominant configurations forming the compound states. 
The compound damping width $\Gamma_\mu$ of interest is
the spreading width of these many-particle many-hole ($n$p-$n$h)
configurations (which we label by $\ket{\mu}$) over 
the compound states $\ket{\alpha}$.

The spreading width $\Gamma_\mu$ is, by definition, the  energy interval
over which the strength of a given  $\left|\mu\right>$ state
is distributed.
The distribution
may formally be represented by the strength function \cite{BM}
$$
S_\mu(E)= \sum_{\alpha} |\bra{\alpha}\hket{\mu}|^2 
\delta(E-(E_{\alpha}-\bar{E}_\mu)),
$$
where $\bra{\alpha} \hket{\mu}$ is the amplitude of the 
$n$p-$n$h $\left|\mu\right>$-state
contained in the compound level $\ket{\alpha}$ of energy
$E_{\alpha}$, while $E$ refers to the energy relative to the 
centroid $\bar{E}_\mu$ of the strength distribution. Calculated examples
of the above function are shown in Fig.1(a). 
It is noted that the 
strength spreads over a limited number of energy
levels, and never shows a smooth profile,
because of the discreteness of the energy levels.
Furthermore, the strength function varies 
strongly from state to state.
A smoother behaviour 
is obtained by taking the average of $S_\mu(E)$ over 
all $\ket{\mu}$ states lying within an 
energy bin and spin interval, trimming the delta functions
with a smoothing function 
(in the present analysis we use a Strutinsky's Gaussian function
with the Laguerre orthogonal polynomial of 10 keV width).
The averaged strength function $\left<S_\mu(E)\right>$ thus 
obtained is shown in Fig.1(b).
It is customary to define the spreading width 
by the FWHM of  $\left<S_\mu(E)\right>$, denoted by   
$\Gamma_\mu^s$, with the label $s$ referring to the 
average strength function.

Another definition of the spreading width is possible,
making use of the autocorrelation function applied to the
strength function $S_\mu(E)$ of individual $n$p-$n$h states. 
The autocorrelation function
$$
C_\mu(e)= \int S_\mu(E+e)S_\mu(E)dE
$$
expresses the probability of pairwise strengths in $S_\mu(E)$ being
located relative to  another at the energy distance $e$.
If the strength function $S_\mu(E)$ were of Breit-Wigner 
shape of width $\Gamma$, the autocorrelation function would also have
a Breit-Wigner shape, displaying twice the width as that
of the original strength functions. 
The autocorrelation function $C_\mu(e)$ has a physical interpretation
as the Fourier transform of the ``survival probability''
$P_\mu(t) = |\left<\mu|\mu(t)\right>|^2$, which measures the
probability of remaining in the state 
$\left|\mu\right>$ during its time evolution 
$\left|\mu(t)\right>=e^{-iHt}\left|\mu\right>$.
For the case of the Breit-Wigner strength function,
$P_\mu(t)$ decays exponentially with a decay constant given by 
$\hbar/\Gamma$.
We average 
$C_\mu(e)$ over many $\left|\mu\right>$ states
in an energy bin and spin interval and make the same smoothing 
as described above for the strength function $S_{\mu}(E)$.
It is remarked that the autocorrelation function $C_\mu(e)$ contains a
delta-function peak at $e=0$ proportional to 
$\sum_{\alpha} |\bra{\alpha} \hket{\mu}|^4$, which
we remove in the
 following analysis, since this peak corresponds to 
the asymptotic value of $P_\mu(t)$ at the
$t\rightarrow \infty$ limit. 
The resultant autocorrelation function 
$\left<C_\mu(e)\right>$ is shown in Fig.1(c).
The {\it correlational spreading width}  can be defined  
as half the value of FHWM of the autocorrelation 
function $\left<C_\mu(e)\right>$.
In order to distinguish from the previous definition 
$\Gamma_\mu^s$ in terms of the averaged strength function, 
we denote this new quantity 
$\Gamma_\mu^{corr}$ making use of the label 'corr'.

The most immediate feature observed in 
the calculated autocorrelation function $\left<C_\mu(e)\right>$ as
compared to the average strength 
function $\left<S_\mu(E)\right>$ is 
its narrower profile.
Correspondingly, the correlational
spreading width $\Gamma_\mu^{corr}= 41$ keV extracted from the 
autocorrelation function shown in  Fig.1(c) is 
about a factor four smaller than 
$\Gamma_\mu^s$.

In order to understand this difference
it is useful to look at the  details of the strength functions 
associated with 'individual' $n$p-$n$h states (cf. Fig.1(a)). 
The strength distribution of individual states is typically clustered 
within a narrower energy  interval than that associated with the average 
strength function $\left<S_\mu(E)\right>$ (cf. e.g. the strength function
associated with the 74-th and 75-th $n$p-$n$h states of angular momentum
and parity $I^\pi = 40^+$). 
Also, the position of the dominant strengths deviates from the centroid
position $(E=0)$ and varies between different $\mu$ configurations. 
This variation results in a broad profile of the average strength 
function $\left<S_\mu(E)\right>$.
In contrast, the width of the individual autocorrelation 
functions  $C_\mu(e)$ reflects the clustering of strengths.
Thus, the averaged autocorrelation 
$\left<C_\mu(e)\right>$ forms a peak around $e=0$ whose width is not
influenced by the energy shift of the dominant 
strength, which only gives rise to wide tails 
stretching out to
large positive and negative energies. 
 Since the energy shift does not imply spreading
nor influence the survival probability, we posit 
that the correlational  width 
$\Gamma_\mu^{corr}$ is more appropriate to characterize the
spreading width than the quantity 
$\Gamma_\mu^s$. 
The difference between
$\Gamma_\mu^{corr}$ and $\Gamma_\mu^s$ decreases gradually 
with increasing excitation energy of the $n$p-$n$h states.
However, 
we find from a calculation using an extended basis of 6000 $n$p-$n$h states
that $\Gamma_\mu^{corr},\Gamma_\mu^s = 133, 305$ keV for the levels
$\#1800-\#2100$ at $I=40,41$  indicating 
that around $U \approx 3$MeV there is 
a difference of about a factor of 2 between these 
two quantities. 
At this energy, while the strength of individual $\ket{\mu}$ states
is spread over several hundreds of levels,
the distribution still displays, in most cases,
a strong clusterization around a few
big peaks, and does not show  a smooth Breit-Wigner distribution. 

Our studies have also shown that the difference found
between $\Gamma_{\mu}^{corr}$ and $\Gamma_{\mu}^s $ is related
to the nature of the two-body residual interaction used in the calculations
(cf. Figure 2).
Replacing the surface delta interaction (SDI) by 
a volume-type
delta force ($V(1,2)=v_{\tau}\delta(\vec x_1 - \vec x_2)$), the ratio 
between $\Gamma_\mu^s$ and $\Gamma_\mu^{corr}$ is as large as 
for the SDI. On the other hand,
using  a random two-body interaction 
for which 
the two-body matrix elements 
$v_{ijkl} = \left<ij|V(1,2)|kl\right>$ are replaced with Gaussian random 
numbers, it is found that the resulting 
$\Gamma_\mu^{corr}$ approximately coincides with $\Gamma_\mu^s$,
irrespective of the average strength of the matrix elements.

Before discussing the physics which is at the basis of these results,
it is reasonable to mention that  the SDI or the delta residual interaction
are a better representation for nuclear
structure calculations at moderate excitation energies above the yrast line,
of the residual interaction acting among nucleons, than that provided by 
a random force.
It is well known that the SDI (or the delta interaction) and
the random interaction differ dramatically in the
statistical distribution of two-body
matrix elements $v_{ijkl}$. In fact, the distribution $P(v_{ijkl})$ for the
SDI, plotted in Fig.3 (and for the delta interaction, not shown here), 
exhibits a strong skewness. In other words,  
it has a significant excess for large matrix elements $|v_{ijkl}| > 60$ keV
compared with a Gaussian distribution having the
same r.m.s value $\sqrt{<v_{ijkl}^2>}=19$ keV. 
In fact, the large matrix elements of the SDI
contribute to the r.m.s. value as much as 
the small ones $|v_{ijkl}| < 60$ keV, 
as seen in the right panel plotting 
$v_{ijkl}^2P(v_{ijkl})$, although large matrix elements
appear quite rarely (only 2\% of the total number of matrix
elements). 
On the other hand, the Gaussian random interaction contains no
such contribution from large matrix elements.
The role of the large (and rare) matrix elements of the SDI can be made even
clearer through a
calculation of $S_{\mu}(E)$ and $C_{\mu}(e)$
carried out with a truncated SDI, where only the  
small matrix elements $|v_{ijkl}| < 60$ keV are kept. This 
truncation has a significant effect on the calculated
average strength function $\left<S_\mu(E)\right>$,
diminishing $\Gamma_\mu^s$ to less than half of its original value. On
the other hand, the
average autocorrelation function $\left<C_\mu(e)\right>$ remains almost
unchanged, keeping the original value of $\Gamma_\mu^{corr}$ (cf. Fig.2).
The large matrix elements of the SDI tend to shift
the energies of the levels, rather than mixing 
the $n$p-$n$h configurations around the energy shell.
As a consequence, they have a strong effect 
on $\Gamma_\mu^s$, but not to $\Gamma_\mu^{corr}$.

Seen from the perspective of gamma decay cascades, the strengths
$S_{\mu}(E)$  and $C_{\mu}(e)$ are {\it zero-step} functions, describing
the coupling of $n$p-$n$h states locally at one value of the angular momentum
$I$.  On the other hand, 
the gamma transitions 
$\left|\alpha(I)\right> \stackrel{E_\gamma}{\longrightarrow} 
\left|\alpha'(I-2)\right>$  
taking place between compound energy levels 
of angular momenta $I$ and $I-2$ are described
by the {\it one-step} E2 strength function $S_\alpha^{(1)}(E_\gamma)$ 
while the consecutive gamma transitions
$\left|\alpha(I)\right> \stackrel{E_{\gamma1}}{\longrightarrow} 
\left|\alpha'(I-2)\right> 
\stackrel{E_{\gamma2}}{\longrightarrow} 
\left|\alpha''(I-4)\right> $ are described by
the {\it two-step} strength functions
 $S_\alpha^{(2)}(E_{\gamma1},E_{\gamma2})$.
Figure 4 shows examples
of these two types of strength functions.
Individual one-step strength
functions $S_\alpha^{(1)}(E_{\gamma})$ display considerable fine structures
(Fig.4(a)) which vary for different initial $\left|\alpha\right>$
states while their average over many states becomes
a rather featureless function(Fig.4(b)), 
from which one can extract only the
rotational damping width $\Gamma_{rot}$.
The two-step function  $S_\alpha^{(2)}(E_{\gamma1},E_{\gamma2})$, 
on the other hand, 
exhibits a two-component structure even after averaging over many
states as shown in Fig.4(c,d) and discussed earlier \cite{Matsuo97,Leoni,scar}.
Projected on the $E_{\gamma1}-E_{\gamma2}$ axis,
the two components are characterized by  wide and narrow widths,
$\Gamma_{wide}$ and $\Gamma_{narrow}$ (cf. Fig.4(d)).
On the basis of our results for the autocorrelation function of the 
zero-step mixing discussed above, we shall show below that
the narrow component in the two-step function can
be given a more precise interpretation as a doorway phenomenon
related to the compound damping width. Thus, the two-step function
carries information on the compound damping width $\Gamma_\mu$
as well as on the rotational damping width $\Gamma_{rot}$.

The admixture of
$n$p-$n$h states $\ket{\mu}$ into each compound state $\ket{\alpha}$
produces strengths $|\bra{\alpha} \hket{\mu}^2|$ which 
fluctuate strongly, even at high excitation energies above the
yrast line ($\approx$ 3 MeV), where their distribution
is expected to approach a Porter-Thomas shape \cite{Matsuo93}.
E2 transitions from a given state $\ket{\alpha}$ at angular momentum
$I$ will single out states $\ket{\alpha'}$ at $I-2$, which
contain strong components of the same $\ket{\mu}$ states
as in $\ket{\alpha}$, and this will  also take place in the
second transition to $I-4$. In this sense, the dominant
components $\ket{\mu(I-2)}$ at the midpoint of the two consecutive 
decay steps act as "doorway states" in the two-step cascade.
If the spreading width $\Gamma_\mu$ of the "doorway states"
is considerably smaller than the rotational damping width
$\Gamma_{rot}$, the E2 strength distribution will exhibit structures
which are associated with the "doorway states" having the rotational
energy correlation, and smeared by $\Gamma_\mu$
in both of the decay steps.
Assuming a Gaussian shape (or a Breit-Wigner) 
for the strength function of the $\ket{\mu}$ states, one finds
$\Gamma_{narrow} = 2 \Gamma_\mu$
(or $2.9 \Gamma_{\mu}$) for the width of the narrow
component. 
On the other hand, the
gamma rays that pass through different $\ket{\mu}$ configurations
in the consecutive steps loose the rotational correlation up
to the energy scale of $\Gamma_{rot}$, contributing to the
wide component, whose width $\Gamma_{wide}$ is thus related
to the rotational damping width as $\Gamma_{wide} \approx 2 \Gamma_{rot}$.
One can
estimate that the intensity $I_{narrow}$ of the narrow component should be
inversely proportional to $n_{door}$, which is
the number of doorway $\ket{\mu}$ states contained in 
a typical compound level $\ket{\alpha}$. In terms of $\Gamma_\mu$ and
the average level spacing $D$, one finds, assuming fluctuations to
have a Porter-Thomas shape, that  
$I_{narrow} =1/n_{door} \approx 2D/\Gamma_\mu$ for
Gaussian, and $\approx D/\Gamma_\mu$ for Breit-Wigner distributions, 
respectively.

As seen in Fig.5, 
the expected relation between the narrow width $\Gamma_{narrow}$
of the two-step function  $S_\alpha^{(2)}(E_{\gamma1},E_{\gamma2})$
and the spreading width $\Gamma_{\mu}$ of the $n$p-$n$h states is verified 
by the numerical calculations. 
The correlational spreading width 
$\Gamma_\mu^{corr}$ exhibits a clear relation to the narrow component
width $\Gamma_{narrow}$ for the different interactions discussed before.
These quantities satisfy
the relation $\Gamma_{narrow} \approx (2-3) \Gamma_{\mu}$ expected
from the above consideration.  Figure 5 indicates 
that the intensity of the narrow component, $I_{narrow}$, also follows
the theoretical expectation. 
The agreement within a factor of two
between calculated and estimated values is
regarded as satisfactory, since
such estimates emphasize the basic physics mechanism, while
effects of coherence between different $\left|\mu\right>$
states are not included.
It is noted that
the spreading width $\Gamma_\mu^s$ extracted from the
average strength function  $\left<S_\mu(E)\right>$ 
does not exhibit
any correlation with $\Gamma_{narrow}$ (cf. Fig. 5).

Experimentally, hints of a two-component structure in the
two-dimensional spectra exist \cite{Leoni}, but they are not easy to
extract from a dominant background of non-consecutive
coincidences. The narrow component occurs in the same
region of energies as that associated with the so called "first ridge", 
which consists of
transitions along unmixed rotational bands. Techniques to
study this narrow component will probably include
analysis of fluctuations \cite{Rev,FAM} and spectra of
dimension higher than two \cite{RPM}.

The numerical calculations were performed at the
Yukawa Institute Computer Facility.
The work is supported by the Grant-in-Aid for 
Scientific Research from the Japan Ministry of Education, Science
and Culture (No. 10640267).

\break

\begin{figure}[t]
\centerline{\psfig{file=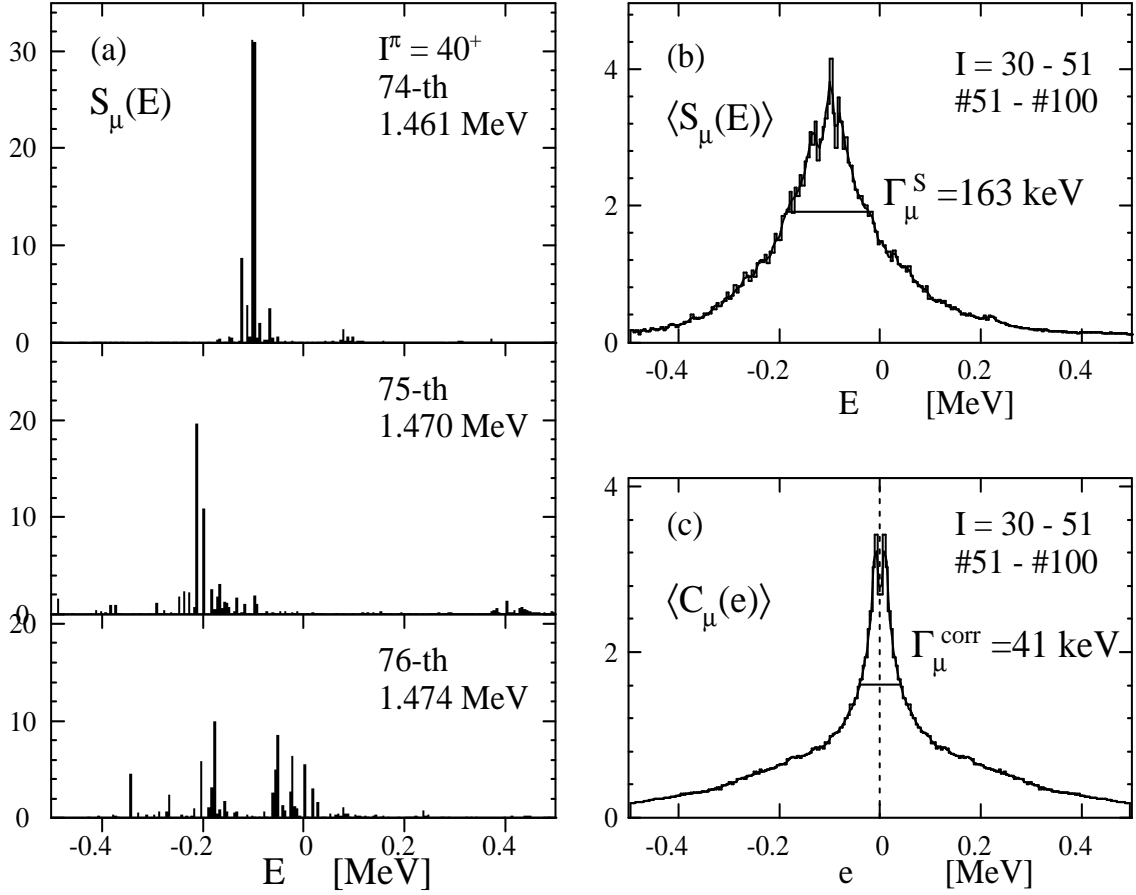,width=15cm}}
\caption{(a) Strength function $S_\mu(E)$ of the individual 
74-th, 75-th and 76-th $\ket{\mu}$ states,  lying around $U=1.47$ MeV
above the yrast line,
at $I^{\pi} = 40^+$.
(b) Strength function $\left<S_\mu(E)\right>$ averaged
over the $\ket{\mu}$ states going from 51-st
to 100-th (for each $I^\pi$), and lying in the  spin interval
$I=30-51$.
(c) Autocorrelation function $\left<C_\mu(e)\right>$,
averaged over the same $\ket{\mu}$ states used in (b).
}
\end{figure}

\begin{figure}[t]
\centerline{\psfig{file=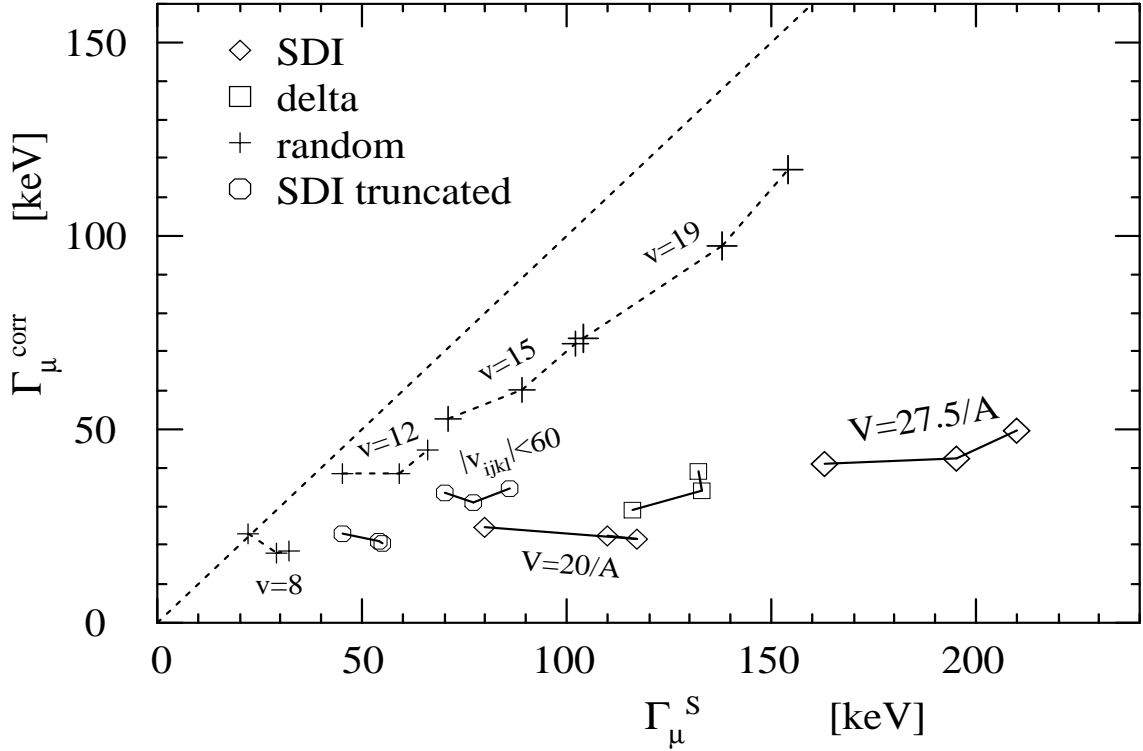,width=15cm,angle=-90}}
\caption{
Comparison of the two spreading widths  $\Gamma_\mu^s$ and $\Gamma_\mu^{corr}$
extracted from the strength function $\left<S_\mu(E)\right>$ 
and from the autocorrelation function $\left<C_\mu(e)\right>$, respectively.
The values shown were obtained averaging 
over the $\ket{\mu}$ states in the spin interval $I=30-51$, 
included in the bins $\#51-\#100$,$\#151-\#200$,
and $\#251-\#300$. We show results obtained
for different interactions:  the SDI with the 
standard strength $V=27.5/A$ MeV 
\protect\cite{Matsuo97} ( symbol $\diamond$) 
and with  
$V=20/A$ MeV ($\diamond$),
the volume-delta force with the strength 
$v_{nn(pp)},v_{np}= 340, 500$ fm${}^3$MeV 
\protect\cite{Matsuo97} ($\Box$), the random two-body interaction
with different r.m.s. values $v=8,12,15,19$ keV ($+$), and the
truncated SDI including only matrix elements 
satisfying $|v_{ijkl}|<40,60$ keV ($\circ$).
}
\end{figure}

\begin{figure}[t]
\centerline{\psfig{file=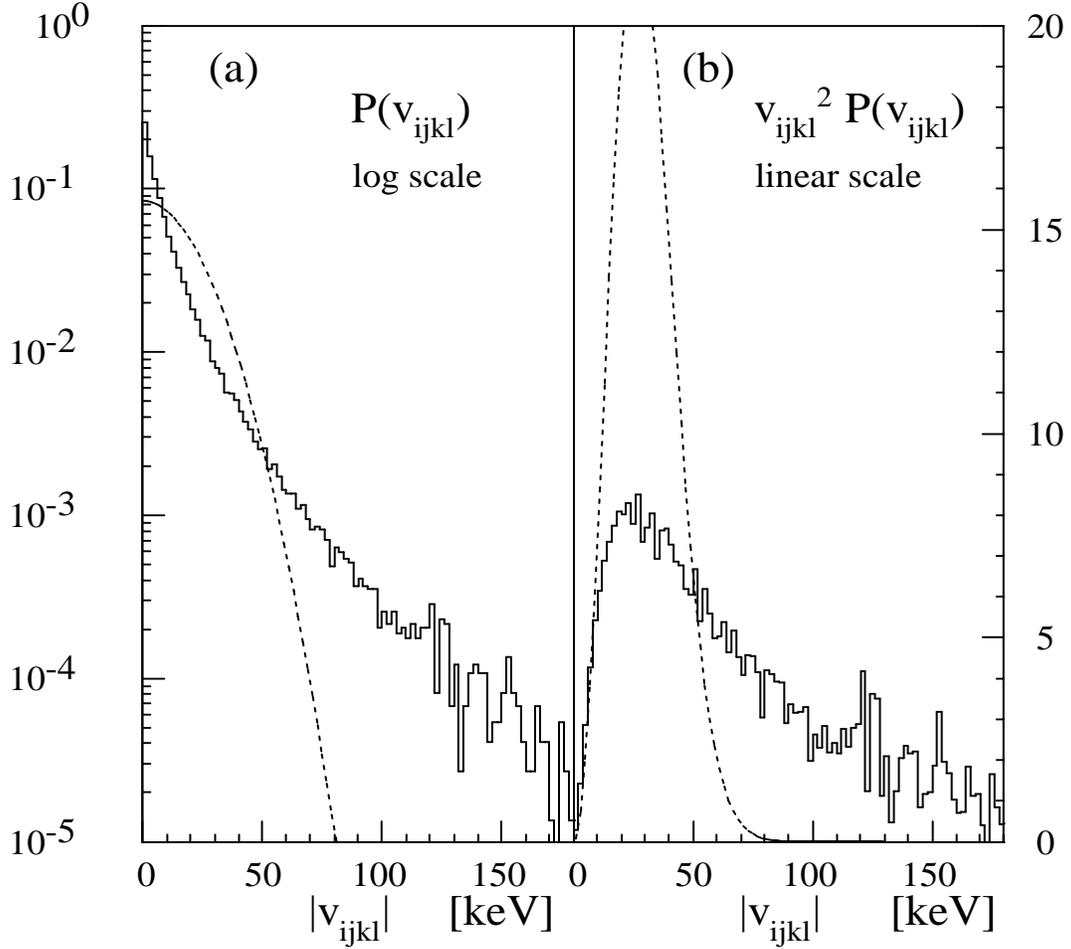,width=14cm,angle=-90}}
\caption{
Statistical distribution of the off-diagonal two-body matrix elements
$v_{ijkl}$ of the SDI, evaluated at $I=40,41$. In 
(a),  the distribution $P(v_{ijkl})$ is plotted. In
(b),  the distribution weighted with $v_{ijkl}^2$ is plotted.
The dotted line
represents a Gaussian distribution whose r.m.s is taken the same
as the SDI (19 keV).
}
\end{figure}

\begin{figure}[t]
\centerline{\psfig{file=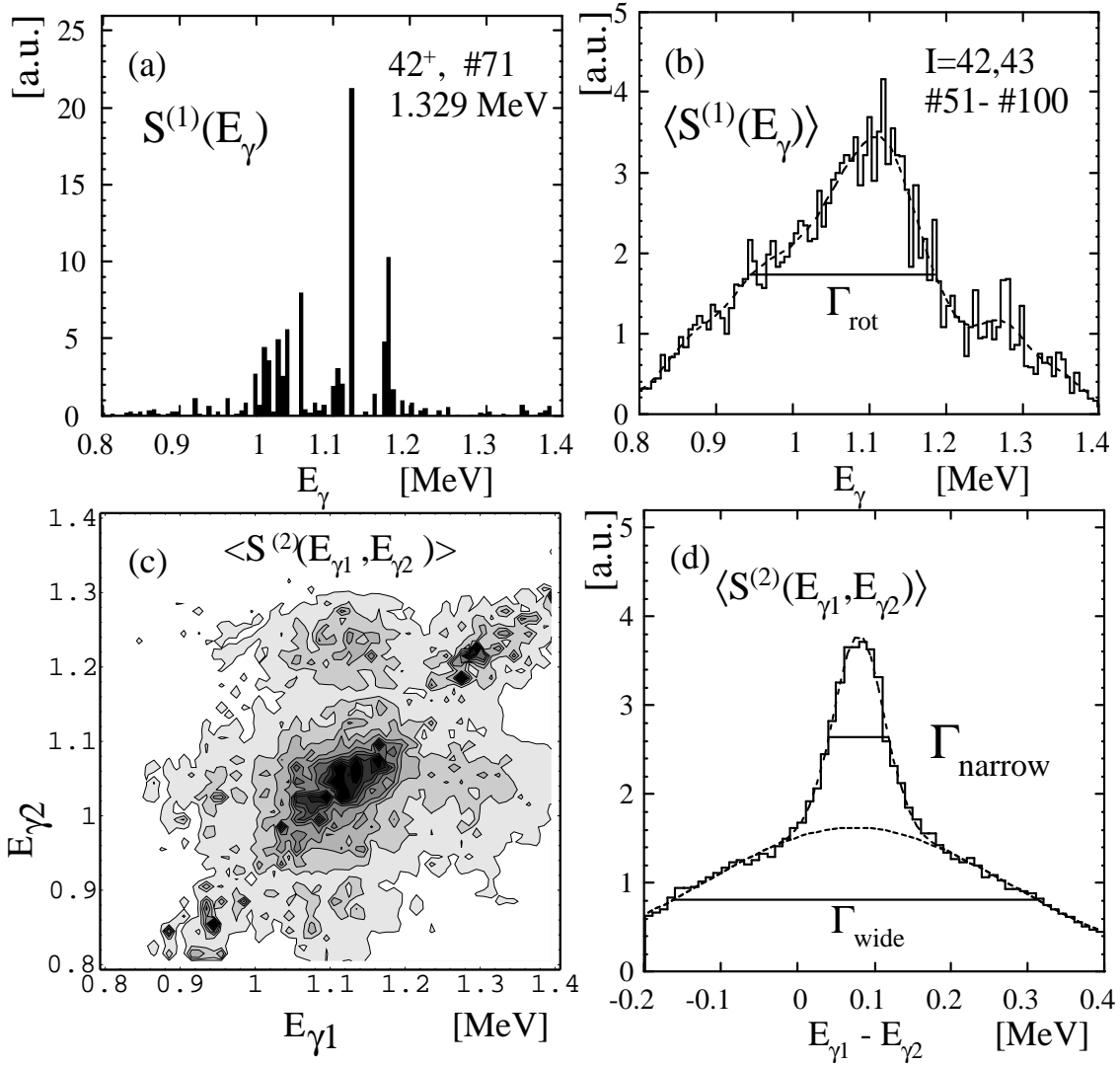,width=15cm}}
\caption{
(a) The calculated one-step strength function
$S_\alpha^{(1)}(E_{\gamma})$ for a typical compound level, the
71-st 42$^+$ states lying at 1.329 MeV above the yrast line.
(b) The average one-step strength function 
$\left<S_\alpha^{(1)}(E_{\gamma})\right>$
for the levels  in the
energy bin including the 51-st to 100th lowest levels (for each
$I^\pi$) at spin $I=42,43$. The associated  width is the 
rotational damping width $\Gamma_{rot}$.
(c) The average two-step strength function
$\left<S_\alpha^{(2)}(E_{\gamma1},E_{\gamma2})\right>$
calculated for the same levels.
(d) Its projection on the 
$E_{\gamma1}-E_{\gamma2}$ axis. The width 
$\Gamma_{narrow}$ and the intensity $I_{narrow}$
of the narrow component are extracted fitting the shape of the 
projection with two
Gaussians.
}
\end{figure}

\begin{figure}[t]
\centerline{\psfig{file=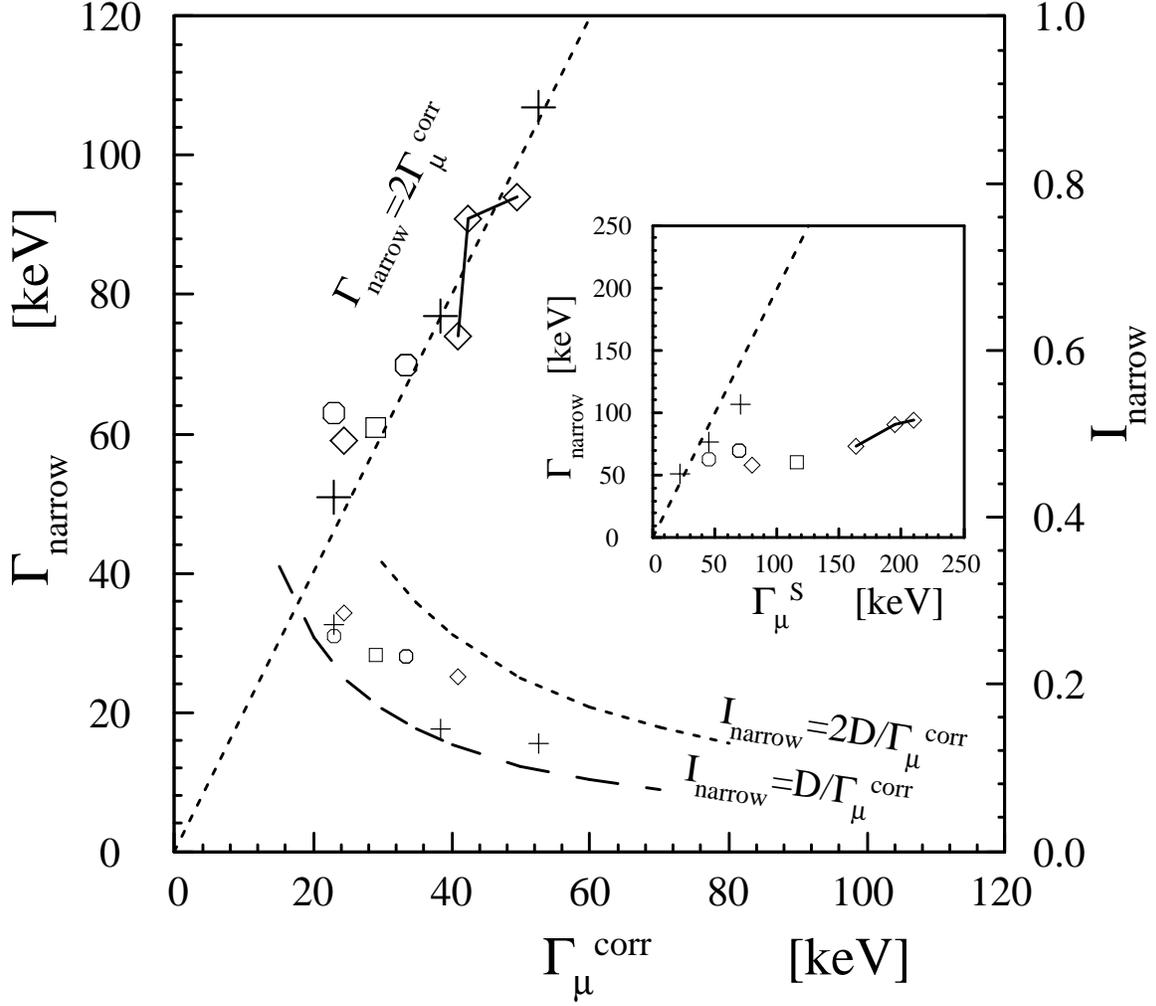,width=15cm}}
\caption{
The values of the width $\Gamma_{narrow}$ (large symbols, left axis)
and of the intensity 
$I_{narrow}$ (small symbols, right axis), associated with 
the narrow component extracted from the two-step strength function 
(cf. Fig. 4(d)), are plotted versus 
the correlational spreading width $\Gamma_\mu^{corr}$
of the $n$p-$n$h states.
Different residual interactions are considered, 
and the averages are taken over the energy bin $\#51-\#100$ and 
over the spin interval
$I=30-51$ (cf. Fig. 2).
In the case of the SDI the results associated with 
 the higher energy bins are also plotted. 
In the inset, the relation between  $\Gamma_{narrow}$ and
$\Gamma_\mu^{s}$ is shown.
}
\end{figure}

\end{document}